\documentclass[aps,twocolumn,prb,preprintnumbers,amsmath,amssymb]{revtex4-2}
\allowdisplaybreaks[4]
\newcommand{\lsim} 
 {\ \raise.35ex\hbox{$<$}\kern-0.75em\lower.5ex\hbox{$\sim$}\ }
\newcommand{\gsim}
 {\ \raise.35ex\hbox{$>$}\kern-0.75em\lower.5ex\hbox{$\sim$}\ }

\usepackage{graphicx}
\usepackage{dcolumn}
\usepackage{bm}
\usepackage{amsmath}
\usepackage{times}
\usepackage[usenames]{color}
\usepackage{natbib}
\usepackage{ulem}

\begin{document}
\title{Anomalous Hall effect in antiferromagnetic perovskites}
\author{Makoto Naka$^{1}$, Yukitoshi Motome$^2$ and Hitoshi Seo$^{3,4}$}
\affiliation{$^1$School of Science and Engineering, Tokyo Denki University, Saitama 350-0394, Japan}
\affiliation{$^2$Department of Applied Physics, The University of Tokyo, Bunkyo, Tokyo 113-8656, Japan}
\affiliation{$^3$Condensed Matter Theory Laboratory, RIKEN, Wako, Saitama 351-0198, Japan}
\affiliation{$^4$Center for Emergent Matter Science (CEMS), RIKEN, Wako, Saitama 351-0198, Japan}
\date{\today}
\begin{abstract}
We theoretically study the anomalous Hall effect (AHE) in perovskites with antiferromagnetic (AFM) orderings. 
By studying the multiorbital Hubbard model for $d$ electrons in perovskite transition metal oxides 
 under the GdFeO$_3$-type distortion within the Hartree-Fock approximation, 
 we investigate the behavior of intrinsic AHE owing to the atomic spin-orbit coupling via the linear response theory. 
We consider the cases where there exist two ($d^2$) and three ($d^3$) electrons in the $t_{2g}$ orbitals,  
 and show that AFM ordered states 
 can exhibit AHE. 
In the $d^2$ case, $C$-type AFM states give rise to dc AHE in metals and optical (finite-$\omega$) AHE in insulators accompanying orbital ordering, 
 while in the $d^3$ case,  a $G$-type AFM insulating state supports the optical AHE. 
By resolving the components in the spin patterns compatible with the space group symmetry, we specify the collinear AFM component to be responsible for the AHE, rather than the small ferromagnetic component. 
We discuss the microscopic origin of the AHE: 
the collinear AFM spin structure produces a nonzero Berry phase from the triangular units of the lattice,  
 activated by the complex orbital mixing terms owing to the GdFeO$_3$-type distortion,  
 and results in the microscopic Lorentz force.
\end{abstract} 



\maketitle
\narrowtext

\section{Introduction\label{sec:intro}}

The anomalous Hall effect (AHE) has extended its platform over the years, 
 from ferromagnets, where it was originally discovered~\cite{hall,karplus}, 
 to other magnetic metals exhibiting different spin structures~\cite{nagaosa,smejkal}. 
Modern research expanded especially after the developments displaying that 
 noncoplanar spin configurations can produce the AHE, 
 by the so-called spin chirality mechanism, associated with the Berry phase felt by conduction electrons~\cite{ye,ohgushi}. 
Such a phenomenon is now called as the topological Hall effect, 
 which is in turn widely used to detect nonplanar spin structures, such as skyrmions~\cite{nagaosa2}. 

On the other hand, coplanar spin structures can also host the AHE under certain conditions, 
 typically discussed for the kagome lattice~\cite{tomizawa,chen}. 
When the orbital degree of freedom is considered, an orbital-driven Berry phase brings about the AHE~\cite{tomizawa}. 
This is indeed detected in a coplanar-type magnetically ordered state in a kagome compound Mn$_3$Sn: 
 a large AHE, compared to that naively expected from the small net magnetic moment from the spin canting, 
 is experimentally observed~\cite{nakatsuji}. 
Engineering large-AHE materials is of interest, 
 and it is discussed that Weyl points in the band structure can be the source~\cite{ito,liu,ikeda,nayak,tanaka}.

Recently, it was shown that certain collinear antiferromagnetic (AFM) states exhibit the AHE as well~\cite{smejkal2, naka}. 
In a theoretical study on an organic system $\kappa$-(BEDT-TTF)$_2${\it X} that shows a collinear-type AFM order, 
 an analytical formula for the AHE is derived by the present authors and coworkers~\cite{naka}.  
We find that the net magnetic moment from the spin canting is not relevant, 
 but the staggered moment, i.e., the AFM order parameter, is essential together with the spin-orbit coupling (SOC).  
In addition, an intuitive understanding was pursued by counting the Berry phase in each triangular unit of the lattice; 
the combination of the collinear AFM order and SOC brings about 
 the microscopic Lorentz force to the conduction electrons, i.e., the origin of AHE. 

All the types of AHE require time reversal symmetry breaking. 
Meanwhile, another spin transport phenomenon under such time reversal symmetry breaking has been discussed: 
 spin current generation, a key ingredient for spintronic devices. 
Similarly to the case of AHE, the expansion of its stage from ferromagnets to other materials is now widely investigated. 
Among them, some AFM systems are pointed out to show spin-split band structures even without a net moment, 
 which result in the spin current generation~\cite{naka2, gonzalez}; 
 conditions for the occurence of such a spin splitting have theoretically been pursued~\cite{hayami, hayami2, yuan, hayami3, yuan2}. 
This mechanism does not require SOC, and the generated spin current shows highly anisotropic behavior depending on the electric field direction. 
Therefore, it is distinct from the spin Hall effect where transverse spin current is driven by the electric field 
 under the spin-momentum locking owing to the SOC~\cite{dyakonov, hirsch, murakami, sinova}. 

In the above-mentioned $\kappa$-(BEDT-TTF)$_2${\it X}, 
 its AFM state in fact hosts such a SOC-free spin current generation, 
 and then, by including the SOC, the AHE is activated~\cite{naka}. 
Here in this work, we show that the AHE appears also in inorganic perovskite-type materials {\it ABX}$_3$ with {\it B} taking transition metals, 
 where we have recently predicted the spin current generation by the mechanism analogous to that in $\kappa$-(BEDT-TTF)$_2${\it X}~\cite{naka3}. 
Indeed, a first-principles band calculation~\cite{solovyev} showed that perovkite transition metal oxides La{\it M}O$_3$ ({\it M} = Cr, Mn, and Fe), in their AFM insulating states,   
 show strong magneto-optical nonreciprocity; 
 this is the finite-$\omega$ counterpart of the AHE in metals that we call the optical AHE in the following, and we expect the symmetry condition should be the same.  

By incorporating the SOC in the framework of the multiorbital Hubbard model for the perovskites, 
 we theoretically investigate the AHE and elucidate its microscopic mechanism. 
In our previous work, the role of the GdFeO$_3$-type distortion from the cubic perovskite was emphasized~\cite{naka3}. 
It gives rise to different sites in the unit cell and crucially affects the spin splitting and the resulting spin current generation. 
Below we will show that the distortion is also essential for the AHE: 
 the orbital mixing effect together with the SOC lead to 
 the microscopic Lorentz force under AFM ordering.  

The rest of the paper is organized as follows. 
In Sec.~\ref{ModelMethods}, we introduce the multiorbital Hubbard model for the $d$ electrons including the atomic SOC 
 in {\it ABX}$_3$; 
 we consider the cases where there exist two ($d^2$) or three ($d^3$)  electrons per {\it B} site.
By applying the Hartree-Fock approximation, we determine the ground state self-consistently, 
 and then calculate the intrinsic AHE by the linear response theory. 
In Sec.~\ref{d2}, the results for the $d^2$ case, where phase competition among different AFM orderings shows up, are shown. 
 The AHE becomes nonzero when the $C$-type AFM ordered metallic states are stabilized; 
 this AFM pattern satisfies the same condition for the spin current generation~\cite{naka3}. 
As for the $d^3$ case shown in Sec.~\ref{d3}, 
 a $G$-type AFM ordered insulating state is stabilized, and results in the optical AHE. 
We discuss the microscopic mechanism for the AHE in Sec.~\ref{mechanism}, 
 and propose material systems to observe our predictions in Sec.~\ref{experiments}. 
Section~\ref{conclusion} is devoted to the conclusion of our work. 

\section{Model and Method}\label{ModelMethods}

\begin{figure}[t]
\begin{center}
\includegraphics[width=1.0\columnwidth, clip]{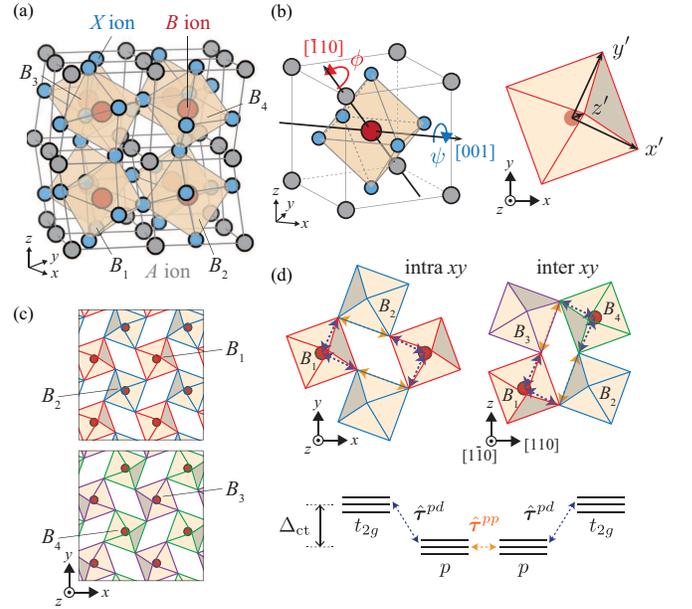}
\end{center}
\caption{
(a) Perovskite structure with the GdFeO$_3$-type distortion. 
{\it B}$_1$-{\it B}$_4$ denote the four {\it BX}$_6$ octahedra in the unit cell, connected by symmetry operations thus crystallographically equivalent. 
(b) Two kinds of {\it BX}$_6$ rotation modes with the angles $\phi$ and $\psi$ in the GdFeO$_3$-type distortion for the {\it B}$_1$ site. 
The right panel shows the $x'y'z'$ axes, the local coordinate defined for each octahedron.
(c) Schematic illustration of the two $xy$ slices in the unit cell.
(d) The real-space hopping paths of the intra- and inter-$xy$-plane NNN bonds between the transition metal {\it B} sites through the ligands {\it X}, 
and their schematic energy diagram.
$\Delta_{\rm ct}$ denotes the $p$-$d$ charge transfer energy.
}
\label{fig1}
\end{figure}

The crystal structure of perovskite {\it ABX}$_3$ with the GdFeO$_3$-type distortion 
 is schematically shown in Fig.~\ref{fig1}(a). 
The global axes, $xyz$, correspond to the crystallographic axes, $abc$, in the space group {\it Pbnm}, 
 or, $cab$, in terms of the equivalent {\it Pnma}. 
There are four {\it BX}$_6$ octahedra in the unit cell, {\it B}$_1$-{\it B}$_4$, 
 while the rotation modes for the GdFeO$_3$-type distortion of {\it B}$_1$ as an example are shown in Fig.~\ref{fig1}(b). 
The two $xy$ slices in the unit cell are shown in Fig.~\ref{fig1}(c). 

Here we extend the multiorital Hubbard model for the transition metal {\it B} sites~\cite{maekawa,mizokawa,mochizuki,mochizuki2}
 constructed in our previous work~\cite{naka3} from the multiorbital $d$-$p$ model. 
We consider the $d^2$ and $d^3$ cases 
 to investigate the behavior of AHE under several AFM spin patterns, 
 therefore restrict ourselves to the threefold $t_{2g}$ orbitals under the octahedral crystalline field. 
These filling factors are chosen referring to previous works: 
 the spin current generation was demonstrated for the $d^2$ case~\cite{naka3}, 
 and the optical AHE was shown in the first-principles band calculation for LaCrO$_3$, a nominally $d^3$ compound~\cite{solovyev}. 
Nevertheless, the resultant conditions for the appearance of AHE should generally be applicable to other filling factors as well. 
%

Our Hamiltonian consists of three parts: 
 ${\cal H} = {\cal H}_{\rm 0} + {\cal H}_{\rm int} + {\cal H}_{\rm SOC}$. 
The first is the tight-binding model, 
\begin{eqnarray} 
{\cal H}_{\rm 0} &=& 
 \sum_{ij \beta \beta' \sigma}^\textrm{NN} [\hat{t}^{dpd}_{ij}(\phi)]_{\beta \beta'} c^{\dagger}_{i \beta \sigma} c_{j \beta' \sigma} \notag \\ 
 &\ & \hspace{3em} + \sum_{ij \beta \beta' \sigma}^\textrm{NNN} [\hat{t}^{dppd}_{ij}(\phi)]_{\beta \beta'} c^{\dagger}_{i \beta \sigma} c_{j \beta' \sigma}, \label{TB}
\end{eqnarray}
where $c_{i \beta \sigma}$ and $c^{\dagger}_{i \beta \sigma}$ are  
 the annihilation and creation operators of a $d$ electron with spin $\sigma$ 
 for the $t_{2g}$ orbital $\beta$ ($=x'y', y'z', z'x'$), respectively, 
 represented in the local $x'y'z'$ axes fixed on the $i$th octahedron as shown in Fig.~\ref{fig1}(b), 
 while the spin axes are globally defined along the crystal axes $xyz$, common to all the sites.
The first and second terms are the nearest-neighbor (NN) and next-nearest-neighbor (NNN) hoppings, 
 considering the second-order and third-order perturbation processes through the ligand {\it X} $p$ orbitals, 
 respectively. 
The $3\times3$ transfer integral matrices, 
 as a function of the major rotation angle $\phi$ of the GdFeO$_3$-type distortion [Fig.~\ref{fig1}(b)], 
 are given as 
\begin{align}
\hat{t}^{dpd}_{ij}(\phi) &= - \frac{1}{\Delta_\textrm{ct}} \ [\hat{\tau}^{pd}_{i;ij}]^\top \  \hat{\cal R}_{ij}(\phi) \ \hat{\tau}^{pd}_{j;ij}, \notag\\
\hat{t}^{dppd}_{ij}(\phi) &= \frac{1}{\Delta_\textrm{ct}^2}\ \sum_k [\hat{\tau}^{pd}_{i;ik}]^\top \  \hat{\cal R}_{ik}(\phi) 
 \ \hat{\tau}^{pp}_{ik;kj} \ \hat{\cal R}_{kj}(\phi) \ \hat{\tau}^{pd}_{j;kj}, \label{hopping}
\end{align}
where $\Delta_{\rm ct}$ is the $p$-$d$ charge transfer energy. 

As introduced in Ref. [\onlinecite{naka3}], 
 $\hat{\tau}^{pd}_{i;ij}$ is the transfer integral matrix from the $t_{2g}$ orbitals of the $i$th {\it B} site  
 to the {\it X} $p$ orbitals shared by the $i$th and $j$th octahedra defined in the local coordinate for the $i$th octahedron.
$\hat{\cal R}_{ij}(\phi)$ is defined by $\hat{\cal R}_{ij}(\phi) = \hat{R}_{i}^\top(\phi) \hat{R}_{j}(\phi)$, 
 where $\hat{R}_i(\phi)$ represents the rotation matrix of the $i$th octahedron, expressed by the Rodrigues rotation formula.
Note that the GdFeO$_3$-type distortion is composed of two rotation modes of the {\it BX}$_6$ octahedra~\cite{okeeffe,naka3}, 
 as shown in Fig.~\ref{fig1}(b); 
 the additional tilting angle $\psi$ is given by a function of $\phi$ as $\psi=\pm\arctan [\sqrt{2}(1-\cos \phi)/(2+\cos \phi)]$.
The third-order perturbation processes are shown in Fig.~\ref{fig1}(d), 
 via two {\it X} sites involving another octahedron labeled $k$ in Eq.~(\ref{hopping}).
$\hat{\tau}^{pp}_{ik;kj}$ is the transfer integral matrix between the two {\it X} $p$ orbitals 
 shared by the NN octahedra pairs ($i, k$) and ($k, j$), 
 defined in the coordinate for the $k$th octahedron; we take the sum over two choices of $k$ giving two paths, 
 as drawn in Fig.~\ref{fig1}(d). 

In the present study, we evaluate $\hat{\tau}^{pd}_{i;ij}$ and $\hat{\tau}^{pp}_{ik;kj}$ using the Slater-Koster parameters, $V_{pd\pi}$,  $V_{pp\sigma}$, and $V_{pp\pi}$. 
Assuming the cubic symmetry in each {\it BX}$_6$ octahedron, the $d$-$p$ hopping between the {\it B} $t_{2g}$ orbital $\beta=\gamma \delta$($=x'y', y'z', z'x'$) and {\it X} $p_{\gamma}$($p_\delta$) orbital in the $\delta$($\gamma$) direction from the {\it B} site is given by $-V_{pd\pi}$; otherwise it is zero because of the orthogonality.
The $p$-$p$ hoppings are classified into three cases, $V_{pp\pi}$, $(V_{pp\sigma} + V_{pp\pi})/2$, and $(-V_{pp\sigma} + V_{pp\pi})/2$, depending on the bond direction and the $p$-orbital configurations~\cite{harrison}.
Note that, owing to the GdFeO$_3$-type distortion, many interorbital matrix elements of the effective hopping between $d$ orbitals through the $p$ orbitals, which are absent for the undostorted case, become nonzero.
We will see below that these orbital mixing terms, as well as the existence of the NNN hoppings  
 that were not considered in our previous study~\cite{naka3}, play crucial roles for the AHE. 

The second part of our Hamiltonian is the onsite Coulomb interactions between the $d$ electrons, introduced in the conventional manner as 
\begin{eqnarray}
{\cal H}_{\rm int} 
&=& U \sum_{i \beta} n_{i \beta \uparrow} n_{i \beta \downarrow} + \frac{U'}{2} \sum_{i \beta \neq \beta'} n_{i \beta} n_{i \beta'} \notag \\
&+& J \sum_{i \beta > \beta' \sigma \sigma'} c^\dagger_{i \beta \sigma} c^\dagger_{i \beta' \sigma'} c_{i \beta \sigma'} c_{i \beta' \sigma} \notag \\
&+& I \sum_{i \beta \neq \beta'} c^\dagger_{i \beta \uparrow} c^\dagger_{i \beta \downarrow} c_{i \beta' \downarrow} c_{i \beta' \uparrow},
\end{eqnarray}
where the number operators are defined as 
 $n_{i \beta \sigma} =c^{\dagger}_{i \beta \sigma}c_{i \beta \sigma}$ and $n_{i \beta} = n_{i \beta \uparrow} +n_{i \beta \downarrow}$. 
$U$ and $U'$ represent the intra- and inter-orbital Coulomb interactions, respectively, $J$ is the Hund coupling, and $I$ is the pair 
hopping interaction. 

The third part is the atomic SOC within the $t_{2g}$ orbitals written as 
\begin{eqnarray}
&{\cal H}_{\rm SOC}&\ = \zeta \sum_i {\bm l}^{\rm loc}_i \cdot {\bm s}^{\rm loc}_i  \notag\\
&=& \zeta \sum_{i \beta \beta' \sigma \sigma'} [ {\bm l}^{\rm loc}_i ]_{\beta \beta'} \ 
  [U_i {\bm s}_i U_i^\dagger]_{\sigma \sigma'} \ c^{\dagger}_{i\beta \sigma} c_{i\beta' \sigma'}, \label{SOC}
\end{eqnarray}
where $ {\bm l}^{\rm loc}_i$ and ${\bm s}^{\rm loc}_i$ are the orbital and spin momentum operators, respectively, in units of $\hbar$ in the local coordinate for the $i$th octahedron; 
 the latter is transformed from the spin in the global axis, ${\bm s}_i = {\bm \sigma}_i / 2$ (${\bm \sigma}_i$: Pauli matrices), 
 used in the electron operators, by the spin rotation operator, 
\begin{equation}
U_i = e^{- i {\bm m}_i \cdot {\bm s}_i \psi_i}  e^{- i {\bm n}_i \cdot {\bm s}_i \phi_i} e^{i \frac{\pi}{4}s^z_i}, 
\end{equation}
with ${\bm m}_i$ and  ${\bm n}_i$ being the rotation axes for the GdFeO$_3$-type distortion: 
for example, ${\bm m}_i$ = $[100]$ and ${\bm n}_i$ =  $[0\bar{1}1]$ for the {\it B}$_1$ octahedron drawn in Fig.~\ref{fig1}(b). 

We treat ${\cal H}_{\rm int} $ within the Hartree-Fock approximation, 
 where the mean fields $\langle c^{\dagger}_{i\beta \sigma} c_{i\beta' \sigma'} \rangle$ are sought for self-consistently in the ground state,  
 assuming four {\it B} sites [{\it B}$_1$-{\it B}$_4$ in Fig.~\ref{fig1}(a)] in the unit cell. 
Since there is no inversion center on either NN or NNN bonds
between the {\it B} sites, 
 the SOC gives rise to spin canting between {\it B}-site spin moments as the Dzyaloshinskii-Moriya interaction works~\cite{dzyaloshinsky, moriya}. 

Using the linear response theory the intrinsic contribution to the Hall conductivity is calculated as 
\begin{align}
\sigma_{\mu \nu}(\omega) = \frac{\hbar}{iN} \sum_{{\bm k}lm} &\frac{f(\epsilon_{{\bm k}l}) - f(\epsilon_{{\bm k}m})}{\epsilon_{{\bm k}l} - \epsilon_{{\bm k}m}} \notag \\
&\times \frac{[J^{\mu}(\bm k)]_{ml} [J^{\nu}(\bm k)]_{lm}}{\hbar \omega + \epsilon_{{\bm k}m} - \epsilon_{{\bm k}l} + i\gamma},
\label{sigma}
\end{align}
where $f(\epsilon_{{\bm k}l})$ is the Fermi distribution function for the Bloch eigenstate with wave vector $\bm k$ and band index $l$.
$[J^{\mu}(\bm k)]_{ml}$ is the matrix element of the $\mu$ component of the total electric current operator between these Bloch eigenstates, $\omega$ is the frequency of the external electric field, and $\gamma$ is the damping factor;  $N$ is the total number of unit cells and 
the lattice constants are set to unity.

In the following, we show results for typical model parameters for the 3$d$ transition metal oxides: 
 $V_{pd\pi}=1$ eV,  $V_{pp\sigma}=-0.6$ eV,  $V_{pp\pi}=0.15$ eV, and $\Delta_{\rm ct}=5$ eV~\cite{mizokawa}. 
As for the Coulomb interaction terms, we adopt the relations $U=U'+2J$ and $J=I$~\cite{sugano}; 
 we vary $U$ while fixing the ratio as $U'=2U/3$ and $J=U/6$, and the rotation angle $\phi$ of the {\it BX}$_6$ octahedra.
The SOC constant, which depends on {\it B}, is also chosen to be a typical value, $\zeta$ =0.04 eV~\cite{sugano} and the damping factor is fixed to $\gamma = 0.01$ eV.


\section{Results}

The SOC specifies the magnetic anisotropy and then the mean-field solutions show certain spin directions. 
In fact, all the possible patterns fall into either of the four types of magnetic structures 
 compatible with the space group ({\it Pbnm}\ /\ {\it Pnma})~\cite{treves,bertaut}. 
For example, when a $C$-type AFM pattern for the $x$ axis spin moment is realized, owing to the symmetry, 
 projections to the other axes must show ferromagnetic ($F$) moments along the $y$ axis and an $A$-type AFM pattern along the $z$ axis; 
 it is written as $C_x F_y A_z$~\cite{solovyev, bertaut2, zhou}. 
For simplification, we will represent each pattern by the major component with the largest projected spin moments, 
 e.g., as $C_x$-type AFM state (see below).
We will introduce others when they appear in the following.  

\subsection{\textit{\textbf{d}}$^2$ system}\label{d2}
\begin{figure}
\begin{center}
\includegraphics[width=0.8\columnwidth, clip]{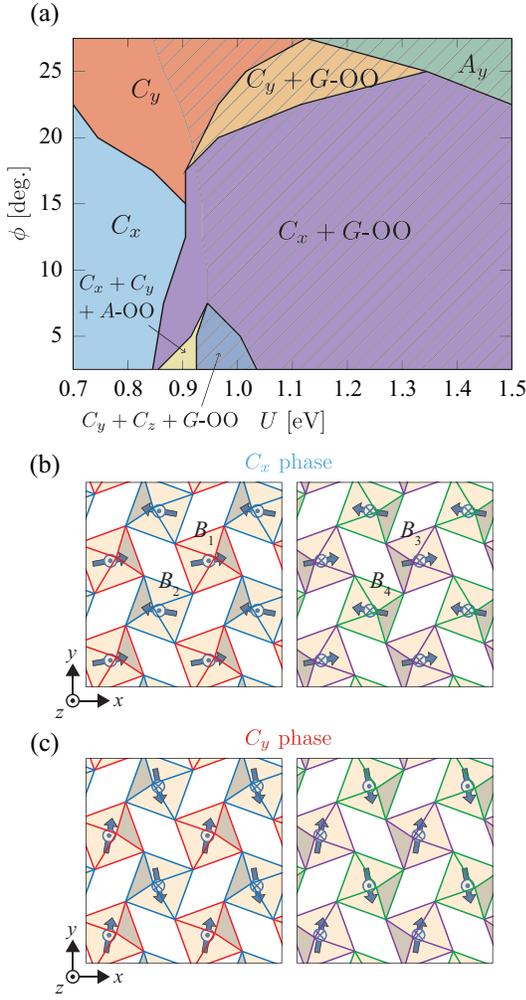}
\end{center}
\caption{
(a) Ground-state phase diagram in the $U$-$\phi$ plane for the $d^2$ case. 
$C_x$, $C_y$, and $A_y$ represent the spatial patterns of the major components of the AFM orders (see text),
 and $A$-OO and $G$-OO denote the orbital orderings.
The hatched area represents the insulating region while the rest is metallic. 
In $C_x + C_y$ and $C_y + C_z$, the two spin patterns coexist in an additive way. 
Schematic illsutrations of the spin structures in the (b) $C_x$ and (c) $C_y$ phases.
}
\label{fig2}
\end{figure}

In the case where there exist two $d$ electrons per site, 
 we have shown in our previous study for the five orbital Hubbard model~\cite{naka3} that, as $U$ is increased, 
 the $C$-type AFM ordering is stabilized in a broad range of $\phi$, 
 where the spin current conductivity becomes nonzero in the metallic region for $\phi \neq 0$; 
 when $U$ is increased further, orbital ordering (OO) sets in and the system becomes insulating. 
Here such overall features are unchanged by restricting to the three $t_{2g}$ orbitals and including the NNN hopping terms and the SOC. 
In the following, we show the ground state properties first (Sec.~\ref{d2:groundstate}) and then the AHE is analyzed (Sec.~\ref{d2:AHE}). 

\subsubsection{Ground state properties}\label{d2:groundstate}

\begin{figure}
\begin{center}
\includegraphics[width=1.0\columnwidth, clip]{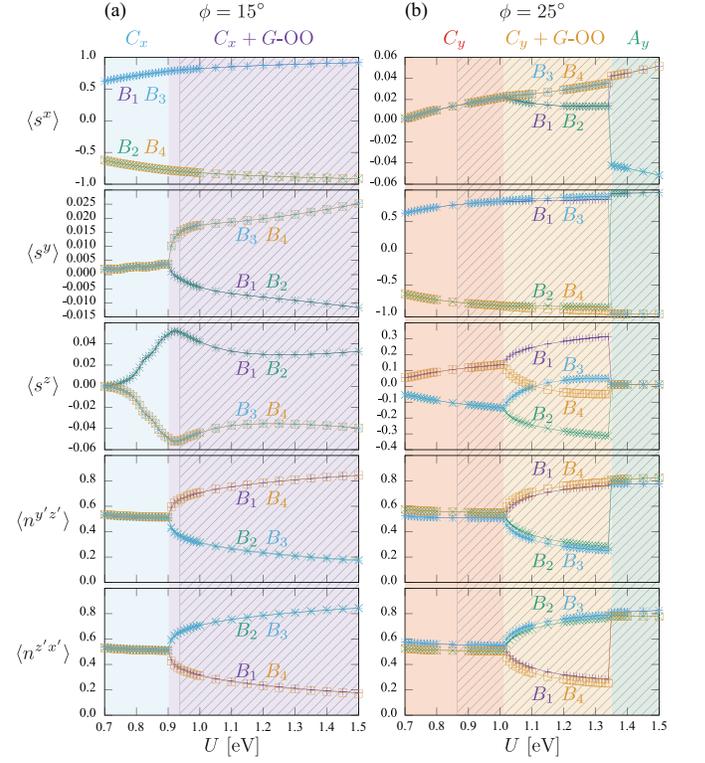}
\end{center}
\caption{
$U$ dependences of the local spin moments along the global crystallographic axis, $\langle s_i^\mu \rangle$ ($\mu=x,y,z$), 
 and the electron densities in the $y'z'$ and $z'x'$ orbitals, $\langle n_i^{y'z'} \rangle$ and  $\langle n_i^{z'x'} \rangle$, 
 for (a) $\phi = 15^\circ$ and (b) $\phi = 25^\circ$. 
}
\label{fig3}
\end{figure}
The ground-state phase diagram on the $U$-$\phi$ plane is shown in Fig.~\ref{fig2}(a). 
We show the region above $U = 0.7$~eV where magnetically ordered states are realized, 
 and for $2.5^\circ \leq \phi \leq 27.5^\circ$ adopted from the realistic range for perovskites; 
 $C$-type AFM ordered states mostly show the lowest energy, 
 except the large-$U$ region for $\phi \gtrsim 25^\circ$ where an $A_y$-type AFM state ($G_x A_y F_z$) is stabilized. 
For $U \gtrsim 0.9$~eV, the $C$-type AFM orders coexist with OO, 
 and the system turns insulating in the hatched area. 
The overall trend is that the $C_x$-type AFM pattern, 
 schematically drawn in Fig.~\ref{fig2}(b), is favored for small $\phi$; 
 on the other hand, $C_y$-type characterized as $F_x C_y G_z$ drawn
 in Fig.~\ref{fig2}(c) has lower energy in the large-$\phi$ region. 
\begin{figure*}
\begin{center}
\includegraphics[width=1.8\columnwidth, clip]{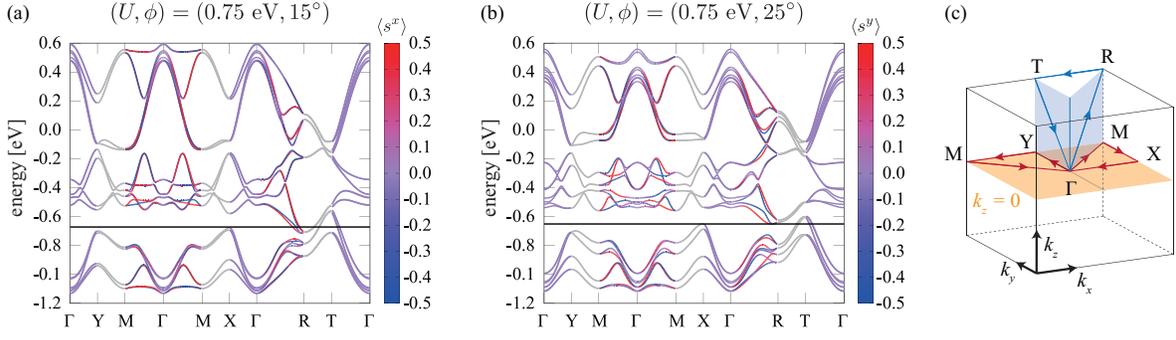}
\end{center}
\caption{
Energy band structures in (a) $C_x$  and (b) $C_y$ metallic phases 
 at $(U, \phi) = (0.75 \ \rm{eV}, 15^\circ)$ and $(U, \phi) = (0.75 \ \rm{eV}, 25^\circ)$, respectively. 
The colors of the bands indicate the magnitude of the expectation value of 
 (a) $\langle s^x \rangle$ and (b) $\langle s^y \rangle$ for their Bloch states. 
The gray lines represent the spin-degenerate bands on the zone boundaries (Y-M, M-X, and R-T) 
 where $\langle s^x \rangle$ and $\langle s^y \rangle$ are not uniquely determined. 
}
\label{fig4}
\end{figure*}

When OO exists, the structural symmetry is lowered, 
 and the magnetic structures do not follow the conditions above. 
In fact, in the small-$\phi$ region,  
 in one phase, an $A$-type OO coexists with the $C_x$-type and $C_y$-type AFM orders, 
 while in another phase, a $G$-type OO coexists with the  $C_y$-type and $C_z$-type ($A_x G_y C_z$) AFM orders. 
Here we will not further discuss their properties, 
 since they only occupy small regions in the phase diagram. 
The former phase actually shows the dc AHE but we leave the survey for future studies. 

In Fig.~\ref{fig3}, 
 mean-field expectation values on the four sites {\it B}$_1$-{\it B}$_4$ are shown 
 as a function of $U$, for $\phi=15^\circ$ and 25$^\circ$.  
The three components of spin moment $\langle {\bm s}_i \rangle$ and the electron densities of the $y'z'$ and $z'x'$ orbitals that show OO are plotted. 
In the $C$-type AFM phases, the $x'y'$ orbitals are almost filled (not shown) to gain the energy of the AFM interaction within the $xy$ plane~\cite{mizokawa}. 
In the $A_y$-type AFM phase, on the contrary, their occupation is small ($\simeq$ 0.4); 
 to gain the AFM interaction along the $z$ direction, 
 the occupations of the $y'z'$ and $z'x'$ orbitals become large [see Fig.~\ref{fig3}(b)]; 
 note that, the orbital occupation pattern here in the $A_y$ phase does not break the {\it Pbnm} / {\it Pnma} symmetry 
 therefore it is not an OO phase. 

For $\phi=15^\circ$ in Fig.~\ref{fig3}(a), the system is in the $C_x$-type AFM phase below $U=0.9$~eV: 
 $\langle s_i^x \rangle$ shows large AFM spin moments with a checkerboard pattern within the $xy$ plane, 
 uniformly stacked along the $z$ direction ($C$-type). 
As for $\langle s_i^z \rangle$, the $A$-type AFM pattern, i.e., ferromagnetic spins within the $xy$ plane 
 stacked in a staggered manner along the $z$ direction, shows up ($A_z$),   
 with spin moments less than one order of magnitude smaller than in $\langle s_i^x \rangle$. 
Finally, $\langle s_i^y \rangle$ is uniformly spin polarized for all sites ($F_y$), 
 with further smaller spin moments: this is the canted AFM component, 
 with a tiny net moment of the order of 10$^{-3}$~$\mu_{\rm B}$ per site. 
Therefore the system shows a nearly collinear AFM state 
 but with small spin canting owing to the SOC. 
 
Above $U=0.9$~eV, the $G$-type OO sets in ($C_x$ + $G$-OO) 
where the electron densities of the $y'z'$ and $z'x'$ orbitals become alternating in the NaCl-type manner. 
There, although the $C_x$ component is almost unchanged from the para-orbital state below $U=0.9$~eV, 
 $\langle s_i^x \rangle$ becomes slightly different among the four sites although it is hardly distinguishable in Fig.~\ref{fig3}(a); 
 $\langle s_i^y \rangle$ also shows small difference between the distinct sites owing to OO, ({\it B}$_1$, {\it B}$_2$) and ({\it B}$_3$, {\it B}$_4$). 
To be precise, on top of the $C_x F_y A_z$ order, another spin pattern $G_x A_y F_z$ is added; 
 namely, it is a phase where  $C_x F_y A_z$ and $G_x A_y F_z$ spin orders coexist with the $G$-OO. 

When $\phi=25^\circ$ [Fig.~\ref{fig3}(b)],  
 the $C_y$-type AFM phase is seen for $U \leq 1$ eV; the largest AFM component is the $C$-type pattern of $\langle s_i^y \rangle$. 
It accompanies $G_z$ and $F_x$: the NaCl-type order in $\langle s_i^z \rangle$ is one order of magnitude smaller than $\langle s_i^y\rangle$, and the canted FM component $\langle s_i^x \rangle$ is of the order of $10^{-2}$ $\mu_{\rm B}$ per site. 
When $G$-OO coexists at larger values of $U$,
 $C_y$ remains to be the major component, 
 while $\langle s_i^y \rangle$ and  $\langle s_i^z \rangle$ split into two components by the OO. 
The added small spin order is characterized as $A_x G_y C_z$: it is an $F_x C_y G_z$ +  $A_x G_y C_z$ + $G$-OO phase. 

Figures~\ref{fig4}(a) and \ref{fig4}(b) show the energy band structures in the metallic state within the $C_x$ and $C_y$ phases, 
 for ($U$, $\phi$) =  (0.75~eV, 15$^\circ$) and  (0.75~eV, 25$^\circ$), respectively. 
The symmetric lines in the first Brillouin zone (BZ) are indicated in Fig.~\ref{fig4}(c). 
The magnitudes of the expectation value of spin moment along the $x$ and $y$ axes, 
 $s^x$ and $s^y$, respectively, for the one-electron Bloch states are indicated as well. 
In both cases, the bands are spin-split in the general $\bm k$ points except for the planes
 $k_x = \pm 0.5$ and $k_y = \pm 0.5$, in units of the reciprocal vectors, at the side edges of the BZ. 
The SOC lifts the degeneracy along $k_x = 0$ and $k_y = 0$ seen in our previous study~\cite{naka3}. 
The Fermi surface is limited around the $R$ and $T$ points,
 basically composed of the $y'z'$ and $z'x'$ orbitals, 
 showing large weight compared to the $x'y'$ orbital~\cite{naka3}. 
We note that the behavior of the spin splitting and spin degeneracy in the $k_x k_y$ plane is 
 analogous to that in $\kappa$-(BEDT-TTF)$_2${\it X}~\cite{naka,naka2,seo}.

\subsubsection{Anomalous Hall effect}\label{d2:AHE}

\begin{figure}[t]
\begin{center}
\includegraphics[width=0.8\columnwidth, clip]{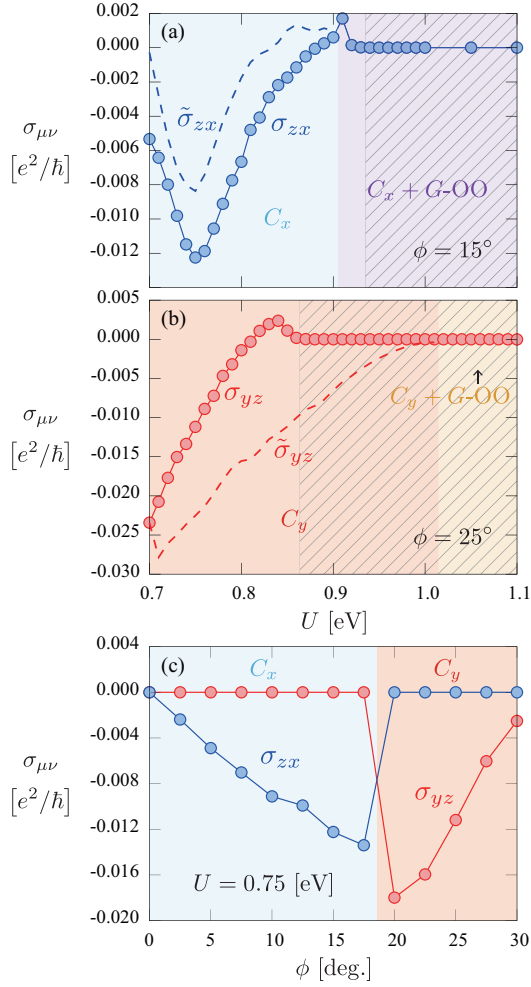}
\end{center}
\caption{$U$ dependence of the dc Hall conductivity $\sigma_{\mu\nu}$: (a) $\sigma_{zx}$ for $\phi=15^\circ$ and 
(b) $\sigma_{yz}$ for $\phi=25^\circ$. 
The hatched area are insulating, where $\sigma_{\mu\nu}=0$. 
 $\tilde{\sigma}_{zx}$ and $\tilde{\sigma}_{yz}$ are the results obtained by 
 artificially restricting the mean-fields to the major collinear AFM orders. 
(c) $\phi$ dependence of them for $U=0.75$~eV.}
\label{fig5}
\end{figure}

\begin{figure}
\begin{center}
\includegraphics[width=0.8\columnwidth, clip]{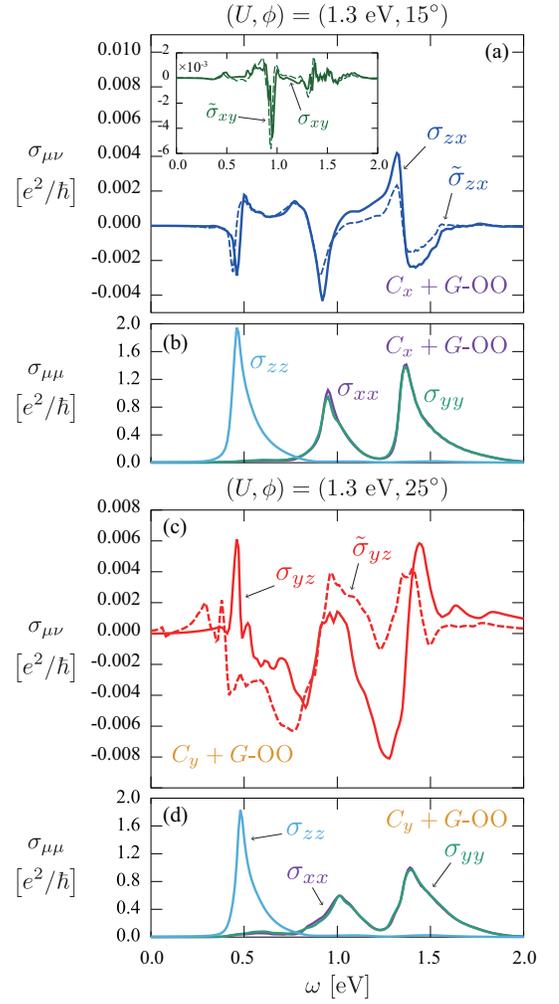}
\end{center}
\caption{
(a) The optical Hall conductivity spectra $\sigma_{zx}(\omega)$ and $\sigma_{xy}(\omega)$ (inset), 
 which are active in the $C_x$ + $G$-OO phase, 
 and (b) the longitudinal conductivity $\sigma_{\mu\mu}(\omega)$ for ($U$, $\phi$) =  (1.3~eV, 15$^\circ$). 
Those for (c) the $C_y$ + $G$-OO phase, $\sigma_{yz}(\omega)$, and (d) $\sigma_{\mu\mu}(\omega)$ for ($U$, $\phi$) =  (1.3~eV, 25$^\circ$). 
$\tilde{\sigma}_{\mu\nu}(\omega)$ are analogous to those in Fig.~\ref{fig5}.
}
\label{fig6}
\end{figure}

The AHE and its finite-$\omega$ counterpart (also called as the magneto-optical Kerr effect), 
 i.e., the dc and optical Hall conductivity, 
for the $d^2$ case are displayed in Figs.~\ref{fig5} and \ref{fig6}, respectively.  
Figures~\ref{fig5}(a) and \ref{fig5}(b) show the $U$ dependences of the Hall conductivity 
 for $\phi=15^\circ$ and 25$^\circ$, corresponding to Figs.~\ref{fig3}(a) and \ref{fig3}(b);
 Figure~\ref{fig5}(c) is their $\phi$ dependence for a fixed value of $U=0.75$~eV. 
The quantities which become active under the magnetic orders are plotted: 
 $\sigma_{zx}$ for the $C_x$-type and $\sigma_{yz}$ for the $C_y$-type AFM patterns. 
In these states, there are ferromagnetic components $F_y$ and $F_x$, respectively, 
 and the AHE appears in the plane perpendicular to them, reminiscent of the case for simple ferromagnets. 
However, their magnitude compared to the net moments are exceptionally large~\cite{nagaosa}.
In fact, when we retain only the major component $C_x$ or $C_y$ 
 by artificially restricting our mean-field solutions to impose the conditions 
 $\langle s_i^y \rangle = \langle s_i^z \rangle =0$ or $\langle s_i^x \rangle = \langle s_i^z \rangle=0$, respectively, 
 the calculated Hall conductivities $\tilde{\sigma}_{zx}$ and $\tilde{\sigma}_{yz}$, also plotted in Figs.~\ref{fig5}(a) and \ref{fig5}(b), 
  roughly recover the original values. 
This indicates that the collinear AFM order is essential for the appearance of the AHE. 

The $U$ dependence of the AHE shows a general trend that 
 it increases as we get away from the insulating state toward small $U$, 
 although the opposite behavior is sometimes seen; 
 for example, a nonmonotonic variation is seen in $\sigma_{zx}$ (and $\tilde{\sigma}_{zx}$) in Fig.~\ref{fig5}(a). 
Such a trend is also seen in the Hubbard model for $\kappa$-(BEDT-TTF)$_2${\it X}~\cite{naka}: 
 the AHE increases when we decrease the Hubbard $U$ in the AFM metallic state, 
 and shows the largest value at the AFM-paramagnetic phase transition boundary. 
This was explained, from the derived analytical formula of the dc Hall conductivity in a single-band model as a limiting case, 
 by its dependence inversely proportional to the AFM order parameter. 
Here also, the major AFM order parameter, i.e., the $C_x$ or $C_y$ component 
 decreases as we decrease $U$, as seen in Fig.~\ref{fig3}. 
The nonmonotonic behavior is owing to the multiband nature of our model here, 
 which give rise to a more complicated band structure (Fig.~\ref{fig4}) than in $\kappa$-(BEDT-TTF)$_2${\it X}. 

One point we emphasize is that the AHE vanishes at $\phi=0$, as shown in Fig.~\ref{fig5}(c). 
This lays out the necessity of the GdFeO$_3$-type distortion; more specifically, it produces orbital mixing in both NN and NNN hopping terms. 
We should also note that the AHE always disappears when we set the NNN terms to zero, even for finite $\phi$. 
These terms are crucial for the AHE, as we will discuss further in Sec.~\ref{mechanism}. 

The optical Hall conductivity is shown for several parameter sets 
 in the AFM insulating states together with the longitudinal optical conductivity, 
  in Fig.~\ref{fig6}. 
The case for the $C_x$ + $G$-OO phase is shown in Fig.~\ref{fig6}(a), 
 where $\sigma_{zx}(\omega)$ is active owing to the $C_x$-type AFM order as discussed above for the dc AHE. 
In addition, $\sigma_{xy}(\omega)$, plotted in the inset, also appears.  
This is because of the lowering of symmetry owing to OO, reflected in the AFM pattern as discussed above: 
 the additional spin component $G_x A_y F_z$ contains the ferromagnetic component in the $z$ direction, 
 therefore consistent with the appearance of $\sigma_{xy}(\omega)$. 
Nevertheless, the optical AHE signal is not much changed when we switch off the $\langle s_i^y \rangle$ and $\langle s_i^z \rangle$ mean-fields and leave only the $\langle s_i^x \rangle$ components ($C_x$ and $G_x$), 
 as shown in $\tilde{\sigma}_{zx}(\omega)$ and $\tilde{\sigma}_{xy}(\omega)$, analogously to the case of dc AHE. 
The values of $\omega$ where the optical Hall signal becomes large roughly correspond to the opitcal $d$-$d$ excitations 
 signalled in the longitudinal optical conductivity $\sigma_{\mu\mu}(\omega)$ [Fig.~\ref{fig6}(b)], 
 although the transverse spectra show more complicated behavior. 
 
In Figs.~\ref{fig6}(c) and \ref{fig6}(d), the optical Hall and longitudinal conductivity spectra for the $C_y$ + $G$-OO phase are shown, respectively.
In this case, only the AHE activated by the $C_y$-AFM order, $\sigma_{yz}(\omega)$, is finite,  
 and there is no additional component. 
This is consistent with the fact that there is no ferromagnetic component additional to $F_y$ under OO, 
 as discussed above: the additional spin pattern is $A_x G_y C_z$ which does not break the time reversal symmetry by itself. 
The spectrum shape of $\sigma_{yz}(\omega)$ is complicated, 
 but the behavior that large values appear at charge transfer peaks in the longitudinal signal holds as well. 

\subsection{\textit{\textbf{d}}$^3$ system}\label{d3}

\begin{figure}
\begin{center}
\includegraphics[width=0.8\columnwidth, clip]{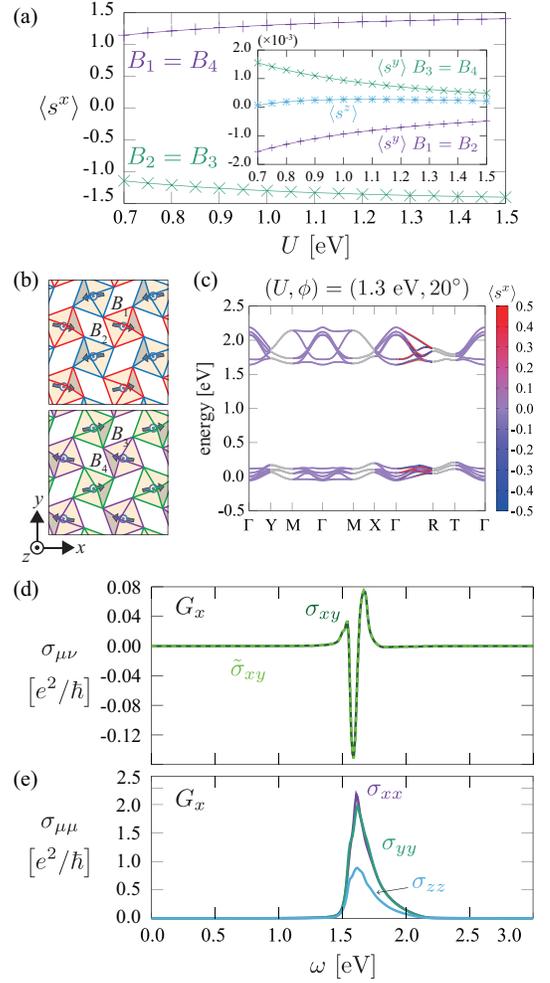}
\end{center}\caption{
(a) $U$ dependences of the local spin moments along the global crystallographic axis, $\langle s_i^x \rangle$, $\langle s_i^y \rangle$ and $\langle s_i^z \rangle$ (inset), for $\phi=20^\circ$. The system is in the $G_x$-type AFM insulating state, whose spin structure is schematically shown in (b).
(c) The energy band structure of the $G_x$-AFM insulator, for ($U$, $\phi$) =  (1.3~eV, 20$^\circ$).
The colors of the bands indicate the magnitude of the expectation value of $\langle s^x \rangle$ for their Bloch states. 
(d) The optical Hall conductivity spectra, $\sigma_{xy}(\omega)$ which is active and $\tilde{\sigma}_{xy}(\omega)$ obtained by artificially extracting the collinear $G_x$-AFM order, and (e) the longitudinal conductivity $\sigma_{\mu\mu}(\omega)$.
}
\label{fig7}
\end{figure}

When there are three $d$ electrons per site, 
 as shown in previous works~\cite{solovyev, mizokawa, zhang}, 
 the $G$-type AFM ordered state is generally stabilized. 
Our calculation shows that the ground state at sufficiently large values of $U$ shows the $G_x$-type AFM order  
 with the full pattern of $G_x A_y F_z$. 
This is seen in the $U$ dependence of the mean-field order parameters shown in Fig.~\ref{fig7}(a) for $\phi=20^\circ$; 
 the AFM pattern is schematically drawn in Fig.~\ref{fig7}(b). 
These results do not change much by varying $\phi$, in contrast to the $d^2$ case. 
The $G$-type AFM state is supported by the AFM interaction that is nearly spatially isotropic 
 owing to the high-spin configuration of $t_{2g}$ electrons, without the orbital degree of freedom.  
In fact, the largest component $\langle s_i^x \rangle$ becomes nearly fully polarized  ($|\langle s_i^x \rangle|=3/2$) at large $U$ as seen in Fig.~\ref{fig7}(a). 
Now the $A_y$ and $F_z$ components from the spin canting are much smaller; the net moment is less than 10$^{-3}$ $\mu_{\rm B}$, as plotted in the inset of Fig.~\ref{fig7}(a).  
Note that the symmetry of this AFM pattern is the same as the $A_y$ phase in the $d^2$ case seen above, 
 whereas the major component is different [see Fig.~\ref{fig3}(b)]. 

The band structure in the $G_x$ phase for ($U$, $\phi$) = (1.3 eV, 20$^\circ$) is shown in Fig.~\ref{fig7}(c). 
One can see a large band gap of about 1.5~eV, in which the Fermi level situates: the system is insulating. 
The spin splitting due to the AFM ordering is seen in the general $\bm k$ points in the BZ, similarly to Fig.~\ref{fig4}; 
however, there is a difference that $\langle s^x \rangle=0$ on the $k_xk_y$ plane at $k_z = 0$.
This is because the sign of the spin splitting is reversed between $k_z > 0$ and $k_z < 0$, in contrast to the case of the $C$-type AFM phases. 
This difference gives the distinct behavior in the spin current generation between the $C$- and $G$-type AFM phases, as discussed in Ref. [\onlinecite{naka3}].
The optical AHE spectrum $\sigma_{xy}(\omega)$ appears at the energy region across this gap, as shown in Fig.~\ref{fig7}(d); 
 the longitudinal optical transitions are also seen there in a nearly isotropic manner [Fig.~\ref{fig7}(e)], 
 consistent with the high-spin $(t_{2g})^3$ configuration. 
Although one naively does not expect the appearance of AHE in the $G$-type AFM order as it is an apparent N\'{e}el state, 
 under the GdFeO$_3$-type distortion 
 the up- and down-spin sites are no longer connected by symmetry operations of time-reversal combined with translation 
 thus results in the nonzero optical AHE. 

\section{Discussions}\label{Discussions}

\subsection{Mechanism of anomalous Hall effect}\label{mechanism}
\begin{figure*}[t]
\begin{center}
\includegraphics[width=1.5\columnwidth, clip]{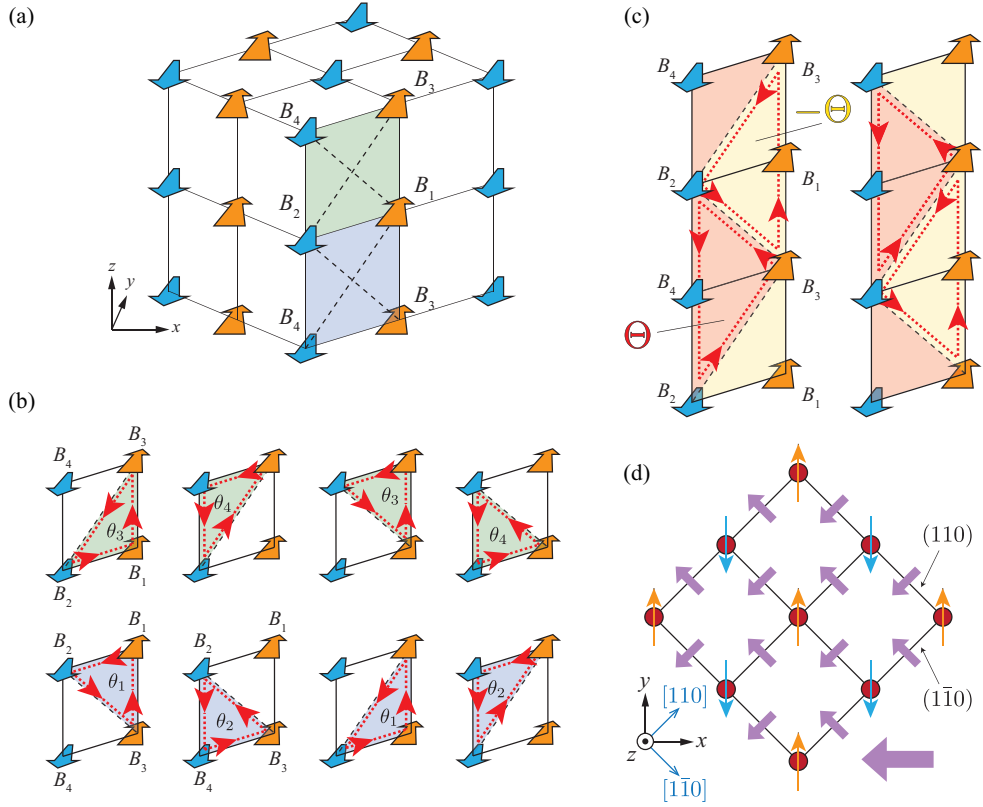}
\end{center}
\caption{(a) Schematic view of the $C_y$-type collinear exchange field. 
The arrows represent the directions of the local exchange field ${\bm h}_i = (0, \pm h, 0)$ at the transition metal {\it B} sites. 
We focus on the highlighed square plaquettes on the $(1\bar{1}0)$ plane, where the broken lines show the NNN bonds.
(b) The smallest triangular paths and fictitious magnetic fluxes. 
The direction of each path is defined to be counterclockwise when viewed from the $x$ axis.
(c) The 
triangular paths composed of pairs of triangles in (b), and fictitious magnetic fluxes acting on up-spin electrons.
The two panels denote the two ways of tilings of the triangular paths.
(d) The top view of the magnetic flux distribution in the $C_y$-type AFM state. 
The small purple arrows represent the fictitious magnetic fluxes penetrating the square plaquettes on the $(1\bar{1}0)$ and $(110)$ planes and the large one is the net flux along the $-x$ direction. 
}
\label{fig8}
\end{figure*}

Now based on the results, 
 we discuss the microscopic mechanism of the AHE presented in this work. 
As shown above, when the spin ordering contains an $F$ component, namely, when the system holds a net moment, 
 the AHE in the plane perpendicular to its direction is activated. 
However, unlike usual ferromagnets where the net moment itself is the source of the Lorentz force 
 to the conduction electrons through the SOC, 
 we have seen that the collinear AFM component is rather essential for the AHE. 
This suggests the existence of a fictitious magnetic field triggered by the AFM order combined with the SOC, which we discuss in the following. 

We purchase a real space picture by taking the $C_y$-type AFM state as an example, since it is realized in several compounds, such as in {\it A}VO$_3$ ({\it A}$=$La-Y)~\cite{miyasaka} and CaCrO$_3$~\cite{komarek}, while other AFM states can also be chosen to follow similar procedures. 
We assume the collinear $C_y$ pattern without other minor components and see how a fictitious magnetic field along the $x$ direction appears, which gives rise to the AHE in the $yz$ plane seen in Sec.~\ref{d2:AHE}.

For this purpose, we analyze a simplified mean-field Hamiltonian: ${\cal H}_{\rm AFM}^{\rm MF} = {\cal H}_{\rm 0} + {\cal H}_{\rm SOC} + {\cal H}_{\rm ex}$.
The first and second terms are identical to Eqs. (\ref{TB}) and (\ref{SOC}), respectively, 
 while the third term provides the $C_y$-type collinear exchange field coupled to the $t_{2g}$ electrons, which is given by
\begin{align}
{\cal H}_{\rm ex} = - \sum_{i \beta \sigma \sigma'} c_{i \beta \sigma}^\dagger {\bm h}_i \cdot [{\bm s}_i]_{\sigma \sigma'} c_{i \beta \sigma'},
\end{align}
where ${\bm h}_i=(0, \pm h, 0)$ is the local field on the $i$th {\it B} site, pointing along the global $y$ axis.
The spatial configuration of ${\bm h}_i$ is schematically shown in Fig.~\ref{fig8}(a). 
Similarly to the case of $\kappa$-(BEDT-TTF)$_2${\it X} in Ref.~\cite{naka2}, in ${\cal H}_{\rm SOC}$, the contribution including the $s^y$ operator (in terms of the global axes), referred to as the $s^y$-term in the following~\cite{ls}, is essential to the AHE, while the other terms coupled to $s^x$ and $s^z$ are irrelevant.
Hence, we calculate the fictitious magnetic field acting on the conduction electrons owing to the $s^y$-term, focusing on the smallest triangular paths composed of two NN and one NNN bonds. 
The triangular paths which can generate the fictitious magnetic field along the $x$ direction are in the $(110)$ and $(1\bar{1}0)$ planes. 
Here we first focus on the $(1\bar{1}0)$ plane highlighted in Fig.~\ref{fig8}(a). 
Figure~\ref{fig8}(b) shows all the triangles on the two different square plaquettes composed of the {\it B} ions in the unit cell. 
The magnetic flux $\theta$ penetrating a closed path $C$ is defined by $\exp(i \theta) = \exp(i \oint_{C} {\bm A} \cdot d{\bm r})$, where $\bm A$ represents the vector potential associated with the path. 
For example, on the triangle composed of {\it B}$_1, $ {\it B}$_2$, and {\it B}$_3$ sites [the lower leftmost panel of Fig.~\ref{fig8}(b)], the flux is given by 
\begin{align}
\theta_1 = \arg{(t_{\textit{B}_1\textit{B}_2} t_{\textit{B}_2 \textit{B}_3} t_{\textit{B}_3 \textit{B}_1})},
\end{align}
where $t_{\textit{B}_1\textit{B}_2}$, $ t_{\textit{B}_2 \textit{B}_3} $, and $ t_{\textit{B}_3 \textit{B}_1}$ are the hopping integrals between the lowest eigenstates of ${\cal H}_{\rm SOC} + {\cal H}_{\rm ex}$ on these sites~\cite{ohgushi,tomizawa}.
Since now $s^y$ is a good quantum number for the conduction electrons, conserved during moving around the path, we can separately treat the magnetic fluxes acting on up-spin ($s^y=1/2$) and down-spin ($s^y=-1/2$) electrons.

The fluxes acting on an up-spin electron moving around the smallest triangles are given by four kinds of values, $\theta_1$-$\theta_4$, as shown in Fig.~\ref{fig8}(b): 
$\theta_i$ satisfies the relation $\sum_{i=1,2,3,4} \theta_i=0$. 
They are classified into two types, depending on the arrangement of the exchange fields: an up-spin electron experiences two (one) parallel and one (two) anti-parallel exchange fields for the $\theta_1$ and $\theta_3$ ($\theta_2$ and $\theta_4$) cases. 
We call the former (latter) triangular paths the low (high) energy paths.
Then we can make two kinds of 
triangles composed of pairs of the low or high energy paths with ($\theta_1$, $\theta_3$) or ($\theta_2$, $\theta_4$), 
 with combined phases  $\theta_1 + \theta_3 = - \Theta$ or $\theta_2 + \theta_4 \equiv \Theta$, respectively.
Figure~\ref{fig8}(d) shows the two possible ways of tiling these 
triangular paths in the $(1\bar{1}0)$ plane featured in Figs.~\ref{fig8}(a) and \ref{fig8}(b).
For instance, in the left panel, the {\it B}$_2${\it B}$_4${\it B}$_2${\it B}$_3$ path, associated with $\Theta$, is made up of the two high energy paths, 
while the {\it B}$_3${\it B}$_1${\it B}$_3${\it B}$_2$ path, penetrated by $-\Theta$, is composed of the low energy paths.
Owing to this energy imbalance, the cancellation of the magnetic fluxes $\pm \Theta$ becomes incomplete and the up-spin electrons feel more $-\Theta$ fluxes penetrating the lower energy paths. 

In this way we can add up all the fluxes contributing to the $x$ direction in the unit cells. 
There is another independent $(1\bar{1}0)$ plane shifted from that discussed above by the vector $(-1/2,1/2, 0)$, 
and we find that it gives the same contribution; the fictitious magnetic field points toward the $[\bar{1}10]$ direction.
On the other hand, as for the above argument to the $(110)$ plane, we find that the field points toward $[\bar{1}\bar{1}0]$. 
Consequently, the resulting net magnetic flux is along the $x$ axis as that along the $y$ axis is cancelled out, as shown in Fig.~\ref{fig8}(e), and 
then the up-spin conduction electrons driven by the electric field in the $yz$ plane drift to the perpendicular direction.

The same argument can be done for the case of a down-spin conduction electron. 
In this case, the signs of the magnetic fluxes are inverted and low and high energy paths are interchanged from the up-spin electron case.
As a result, all the low energy paths are penetrated by $-\Theta$, also for the down-spin electron case. 
Therefore, the down-spin electrons drift to the same direction with the up-spin electrons under the electric field, which results in the Hall conductivity $\sigma_{yz}$.

Let us comment on other contributions to the AHE. 
All the AFM patterns discussed in this paper have minor spin components, 
 namely, the spin structures show spin canting owing to the SOC under the distorted structure. 
Then if we consider the triangular paths as above, the spin directions deviate from collinear, and then the spin scalar chirality, which is defined by the triple product of three spins on the triangle, becomes nonzero. 
Therefore, in addition to the Lorentz force from the collinear-type AFM spin component as derived above, 
 there should be a contribution from the spin chirality mechanism to the full AHE.  
This can be one of the reasons for the difference in the calculated $\sigma_{\mu\nu}$ and $\tilde{\sigma}_{\mu\nu}$ 
 shown in the previous section, since the latter extracts the collinear component. 
Another possible reason for the difference is in the band structures for the two cases because of the different mean-field values. 
In fact, our calculations show notable change in the energy bands near the Fermi surface by switching on/off the minor spin components. 
Nevertheless, the $U$ and $\omega$ dependences of the Hall conductivities are well reproduced and the mechanism is inituitively understood from the real-space fictitious field by the contributions from the major AFM components, suggesting that the present AHE is dominated by the collinear AFM order.

\subsection{Relevance to experiments}\label{experiments}

As mentioned Sec.~\ref{d2}, the AFM spin orderings in the GdFeO$_3$-type distorted perovskite 
 with the {\it Pbnm} / {\it Pnma} space group fall into either of the four patterns: 
 $C_x F_y A_z$, $F_x C_y G_z$,  $G_x A_y F_z$, and $A_x G_y C_z$. 
When OO coexists, it breaks the symmetry; then mixing between two of the patterns are seen in our calculations. 
In the following we list some of the candidate materials: AFM metals for observing the dc AHE,  
 and AFM insulators for the optical AHE. 

We suggested in Ref.~\cite{naka3} that $d^2$ compounds with $C$-type AFM ordering 
 might exhibit spin current generation; 
 they are also AHE-active as we have shown above. 
Typical examples are vanadates with a trivalent $A^{3+} $, which are mostly AFM ordered insulators at low temperatures. 
Systematic studies show that a competition occurs between two phases with coexisting AFM order and OO~\cite{miyasaka2}. 
One is typically realized in LaVO$_3$ below $T=$ 140~K, 
 $C_y$-type AFM + $G$-OO~\cite{miyasaka,motome}, reproduced in our phase diagram in Fig.~\ref{fig2}(a). 
Another is seen, e.g., in YVO$_3$ below $T=77$~K, $G_z$-type AFM + $C$-OO~\cite{kawano}, 
 which is not stabilized in our results; 
 this is presumably owing to the role of Jahn-Teller distortions~\cite{mizokawa2}, not considered in our model.  
Investigating the optical AHE across the two AFM patterns in {\it A}VO$_3$ ({\it A}$=$La-Y) should be interesting. 
Chromates with divalent $A^{2+} $ are also nominally $d^2$ compounds, however, not many studies have been conducted so far. 
A rare example is CaCrO$_3$, an AFM metal below $T=90$~K, where experiments indicate the $C_y$-type pattern without OO~\cite{komarek}. 
This corresponds to the $C_y$ metallic phase in our results therefore it is promising to observe the dc AHE $\sigma_{yz}$. 

As for $d^3$ systems, 
 a first-principles band calculation demonstrated nonzero optical AHE in LaCrO$_3$ for the AFM patterns above~\cite{solovyev},  
 while a later experiment suggested its pattern to be either $F_x C_y G_z$ or $G_x A_y F_z$~\cite{zhou}. 
Manganites with divalent {\it A}$^{2+} $, such as CaMnO$_3$ and its lightly electron-doped compounds are also candidates~\cite{koehler}: 
the major AFM component in Ca$_{1-x}$La$_x$MnO$_3$ for $0 \leq x < 0.07$ is assigned to be $G_x$~\cite{ling}, and it is $G_z$ in Ca$_{1-x}$Ce$_x$MnO$_3$ for $0.025 \leq x < 0.075$~\cite{caspi}.
A first-princples band calculation also supports a canted $G$-type ground state in CaMnO$_3$~\cite{ohnishi}.

As in CaMnO$_3$ with La and Ce substitutions, in many perovskites, chemical doping induces carriers with non-integer $d$ electrons per {\it B} site, 
and sometimes results in an AFM metal out of insulating mother compounds~\cite{maekawa}. 
 $e_g$ systems with $3d$ electrons more than $d^3$ are also candidates. 
 Moreover, $4d$ and $5d$ perovskites, incorporating stronger SOC than $3d$ systems~\cite{calder, du, cui}, may exhibit large AHE when AFM orders are realized. 
We leave the calculations using parameters suitable for $4d$ and $5d$ electrons, and for other filling factors than $d^2$ and $d^3$ for future issues. 
Finally, first-principles band calculations should provide quantitative estimation of the AHE and complementary information about the material-specific electronic structure, 
 compared to our model study which can expand the parameter range and enables us to obtain systematic views. 

\section{Conclusion}\label{conclusion}
We have proposed the appearance of the dc and optical AHE in AFM perovskites with $d^2$ and $d^3$ electron configurations.
The AHE originates not from the net magnetization by spin canting but from the collinear AFM ordering, 
in contrast to the conventional AHE in ferromangets and the topological Hall effect due to the spin chirality mechanism in noncoplanar magnets. 
The microscopic origin is the cooperative effect of the fictitious magnetic field, emerging from the synergy among the atomic SOC, the interorbital hoppings due to the GdFeO$_3$-type distorion, and the local exchange field owing to the collinear AFM ordering; these give rise to the net Lorenz force acting on the conduction electrons.
This mechanism is not limited to $d^2$ and $d^3$ compounds, e.g., vanadates and chromates, but applicable to a wide variety of perovskites showing AFM ordering.

\begin{acknowledgments}
The authors would like to thank H. Kishida, J. Matsuno, I. V. Solovyev, and K. Yamauchi for valuable comments and discussions. 
This work is supported by Grant-in-Aid for Scientific Research, No. 19K03723, 19K21860, and 20H04463, and the GIMRT Program of the Institute for Materials Research, Tohoku University, No. 202112-RDKGE-0019.
\end{acknowledgments}





\begin{thebibliography}{99}
\bibitem{ls}
Note that, in ${\cal H}_{\rm SOC}$, the orbital angular momentum $ {\bm l}^{\rm loc}_i$ is defined in the local coordinates, $x'y'z'$, following the $t_{2g}$ orbitals defined there. 
Then the $s^y$-term is not simply written as $l^\mu s^y$, where $l^y$ is the angular momentum along the global $y$ axis, because the high-energy $e_g$ orbitals are not taken into account.
\end{thebibliography}
\end{document}